# Analysis of Potential Biases and Validity of Studies Using Multiverse Approaches to Assess the Impacts of Government Responses to Epidemics

Jeremy D. Goldhaber-Fiebert, PhD

Stanford Health Policy

Department of Health Policy, Stanford School of Medicine

Center for Health Policy, Freeman Spogli Institute

Stanford University

Stanford, CA, USA 94305

jeremygf@stanford.edu

Abstract:

We analyze the methodological approach and validity of interpretation of using national-level time-series regression analyses relating epidemic outcomes to policies that estimate many models involving permutations of analytic choices (i.e., a "multiverse" approach). Specifically, we evaluate the possible biases and pitfalls of interpretation of a multiverse approach to the context of government responses to epidemics using the example of COVID-19 and a recently published peer-reviewed paper by Bendavid and Patel (2024) along with the subsequent commentary that the two authors published discussing and interpreting the implications of their work. While we identify multiple potential errors and sources of biases in how the specific analyses were undertaken that are also relevant for other studies employing similar approaches, our most important finding involves constructing a counterexample showing that causal model specification-agnostic multiverse analyses can be incorrectly used to suggest that no consistent effect can be discovered in data especially in cases where most specifications estimated with the data are far from causally valid. Finally, we suggest an alternative approach involving hypothesis-drive specifications that explicitly account for infectiousness across jurisdictions in the analysis as well as the interconnected ways that policies and behavioral responses may evolve within and across these jurisdictions.

*Introduction*

There is a desire to analyze how policies impact epidemic outcomes, especially in light of the global COVID-19 pandemic that began in 2020. In most part this desire is motivated by causal questions regarding what would have happened if alternative policies had been implemented; what would have happened if policies had been implemented with different intensities and/or timing; or, in a more forward looking way, what will happen if similar policies were used in the future.

To the extent that studies examine such questions, we evaluate the methodological approach and validity of interpretation of using national-level time-series regression analyses relating epidemic outcomes to policies that estimate many models involving permutations of analytic choices (i.e., "multiverse" approaches). Specifically, we examine a recently published paper by Bendavid and Patel entitled "Epidemic outcomes following government responses to COVID-19: Insights from nearly 100,000 models" in *Science Advances*.(1) To fully understand the implications of this work, our examination also includes a commentary on this published work that the authors subsequently published entitled "100,000 models show that not much was learned about stopping the COVID-19 pandemic" in STATNews.(2)

Our approach consists of two parts. The first involves considering potential sources of bias in the analyses as undertaken in the published work. The second involves considering the validity of how the analysis and its findings are interpreted. Finally, given that we identify multiple potential errors and biases and challenges to the validity of interpretation of the approach, we briefly suggest an alternative approach involving hypothesis-drive specifications that explicitly account for infectiousness across jurisdictions in the analysis as well as the interconnected ways that policies and behavioral responses may evolve across these jurisdictions.

*Approach and Findings*

**Error in Reported Results?** The analysis by Bendavid and Patel makes approximately 3,000 of its approximately 100,000 estimates using the policy variable (Vaccine Policy [specifically vaccine policy related to COVID per the Oxford source used (3), which the authors term "vaccine availability"]) and the time periods of early 2020 or all of 2020.[1] Until the very end of 2020 no vaccine for COVID was available in any jurisdiction.[2] Therefore, there should be no variation in the policy variable and no estimates should be possible. What is perhaps more concerning is that in these approximately 3,000 estimates, the large majority show a policy coefficient of "harmful" (79%) suggesting the potential for underlying biases in the regression machinery used by the authors. The inclusion of the data points is relevant to multiple figures and papers in the main paper as well as counts and proportions reported in the text itself. These results are also made available for exploration by the authors in their Shiny app (4).

**Is the Analysis Meant to Be Interpreted Causally or Not?** In their peer reviewed publication, the authors are careful to note that they are estimating associations and cannot comment on the counterfactual of what would have happened if policies had not been implemented as observed. While one might argue that the authors' choice of language regarding policies being "helpful" or "harmful" might belie disavowal of causal interpretation, what is concerning is that the authors themselves write in their STATNews commentary on their work:

> *"We modeled the policy effects in nearly 100,000 different ways, representing nearly 100,000 theories, each a flavor of a question about the effects of government responses to Covid-19. 'Did stay-at-home policies flatten the curve?' or 'Did closing schools decrease the spread of infections?' were among the hypotheses we tested."*

Their description of their own work in STATNews is explicitly causal, stated in the language of "effects" of policies. Given the statements made in the peer-reviewed work disavowing causal estimation, it is incumbent on the authors to correct/retract the causal description of what the work is doing that appears in the STATNews commentary and to reaffirm that they do not believe that the estimates from their published research paper are causally valid.

---

[1] An additional approximately 1,700 results use Vaccine Policy as the policy variable and are estimated for all of 2020 and 2021. While estimable, given that for all countries in the analysis, the 2020 time period has virtually no variation and there appears to be bias in the estimates for this period, it is possible that these estimates also contain some bias. However, this would require closer assessment than is possible with what has been published/made publicly available.

[2] According to World Health Organization data aggregated on the Our World In Data website (https://ourworldindata.org/grapher/cumulative-covid-vaccinations; accessed September 5, 2024), the number of countries with any delivered doses of vaccines and number of delivered doses of vaccine were exceedingly small as of December 18, 2020 (the last day for which the shortest lag used in the multiverse analysis would still result in an outcome for the All of 2020 analysis). Only China and the US had delivered >500,000 doses (only 4 countries: China, US, Russia, and Canada had exceeded 10,000 doses delivered by that date).

**Overinterpretation of Results and Implications?** The authors' multiverse study design intends to remain neutral about the correct causal model specification as well as a number of other potential analytic choices like time-period of analysis, which policy characterization to use, what outcome, what time-lag, whether to include country-level fixed effects, etc. The authors' process involved running regressions with these many combinations of specifications and other analytic choices and then reporting that the relationship between the policy intensity and outcomes was rarely significant, could be either positive or negative, and did not show systematic patterns of when it was positive/negative and/or significant/non-significant based on the various analytic choices. Based on this, they conclude from this that nothing can be learned about the relationship of policies and epidemic outcomes with a high degree of certainty, responding to the fact that various prior publications had done specific analyses and reported strongly helpful or strongly harmful effects of policies on outcomes.

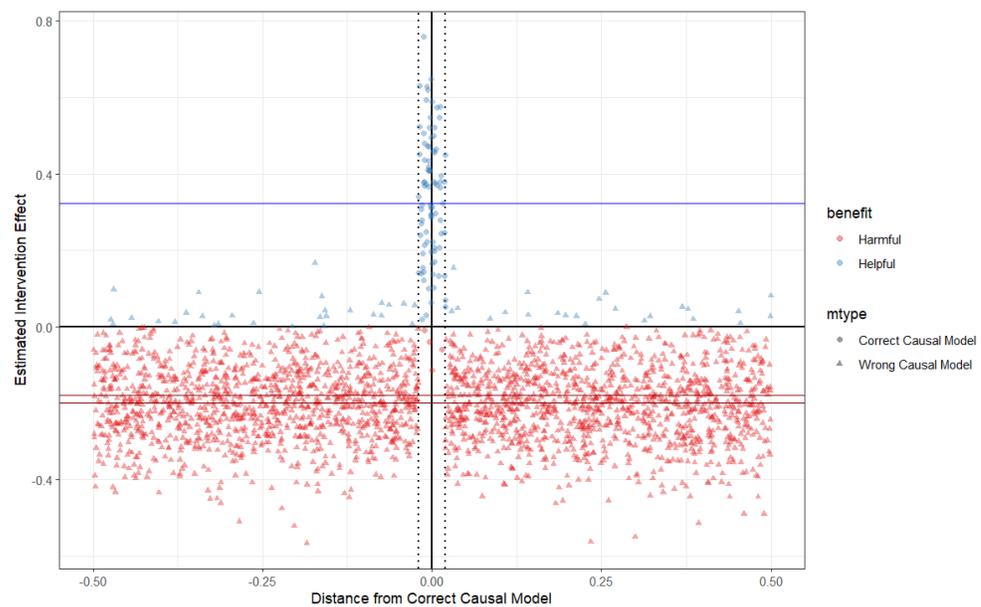

But I would argue that the authors' conclusions are an overinterpretation of their methods and of their results and the implications. My argument is as follows. There is some underlying true causal data generating process for the epidemic outcomes. We do not know what this true causal process is. We also do not know whether policies have causal effects in this process, and likewise, we do not know the direction/magnitude of such effects if they truly exist.

Having observed the realization of the data that we have, we could specify many models whose parameters we could then estimate with regressions. Some of our specifications might be the same as the true underlying data generating process (or very close to it) and some may be far from it and highly confounded. We might even reasonably suppose that most models we specify are not going to be close to the true causal model.

If we adopted a multiverse approach and estimated the parameters for all of the models (the causally valid models as well as the confounded and not valid models), we might produce a figure like the one shown above. In it, we see that most models show a harmful effect of the policy (negative coefficient estimated), but we also see that most models are not the correct/true causal model. The correct causal model estimates a helpful effect of the policy.

If our interpretation of our results involved just looking at the proportion of models that show a helpful (harmful) effect (as the authors do in their paper), we see that it is very small (i.e., a low percentage of the estimated policy effects are "helpful"). So, is this evidence that the policy is (most likely) harmful? I would argue it most certainly is not because having *many* biased estimates of an effect does not add to one's knowledge about the true unbiased effect. The question is: what is the estimated effect or range of effects under causally valid specifications?

And that is why it is essential to conduct hypothesis-drive research. To conduct such research, I must lay my cards on the table and state and justify why I believe the model that I am specifying and then estimating is the causally valid one. What are the biological, mechanistic, behavioral, and other factors that other evidence and theory suggest are at play? Having done this and formalizing it into a causal model, one can then estimate the effect size, direction, and significance **under that model** using data. Data alone will not tell me this. Data will only allow me to make estimates under the specified model.

**Unique Outcomes Overstated?** If one had a single dataset that had people's measured height and some predictor of heigh, one would not run two regressions with using this dataset with the outcome being height in meters and the other outcome being height in feet and then claim that two independent assessments were done, it is not clear that the 9 outcomes reported in the paper are truly independent from each other (i.e., there are many fewer than 100,000 different ways of analyzing the data as the authors title their work). Really there are 3 main outcomes: detected cases, deaths, and infections. While several entities report time series for cases and deaths, the underlying data used by the entities to construct these time series is largely similar (e.g., derived from the same county, state, or national reporting systems). Even more so, the estimates of infections produced by IHME are mechanically based on deaths data, data on age distributions, and assumptions about age-specific case fatality rates and hence are definitely not independent from the IHME deaths outcome (5). Furthermore, as there are more of some types of outcomes (3 cases series) than others (2 deaths series) but the analysis essentially treats all outcomes as the same, this apparent multiplicity of outcomes both inflates the appearance of independent tests and also idiosyncratically reweights the findings reported in the paper towards outcomes with more reported series.

**Bias Due to Level of Aggregation?** The analysis is conducted at the national level, with national-level estimates of the intensity of policies used as independent variables. However, policies were promulgated and implemented heterogeneously within many countries, especially large countries (e.g., we have seen substantial heterogeneity of policy timing and intensity in our own work cataloging the evolution of pandemic policies in California's counties (6)). Essentially the analysis aggregates over these differences and takes the average. As a stylized example, if half the states in a country did not implement a policy and the other half did, the policy level is set to 50% and used as the predictor in the regression. But given that epidemics evolve as non-linear systems, it seems plausible that this approach suffers from bias due to Jensen's Inequality, which states that the value of a non-linear function computed at its inputs' expectations will not necessarily be the same as computing the expectation of the non-linear function computed over samples from the distribution of its inputs. In other words, computing functions of averages can induce bias.

The bias due to aggregation can occur for both dependent and independent variables in the regressions run by the authors. If we take the United States as an example, across the 50 states, incidence in their initial epidemics peaked at various points over a 10-week period and had rapid rates of decline after peaking (7). Aggregating to the U.S. national level would result in a time series for cases with a much longer but lower incidence peak that would resemble none of the epidemic curves for any of the 50 individual states. Likewise, as a policy example, by 2022, U.S. counties differed in terms of percentage of fully vaccinated individuals from >70% to <30% (8). Averaging over these counties, even accounting for population size, would produce a national-level aggregate estimate that is far too low for some counties, far too high for others, and systematically over/underestimated by region.

While the authors did not develop the policy time series that they used, they did make the analytic choice to use it despite the shortcomings it has. It is not clear exactly how to assess the directional bias that may induced by its use, but even if there is not such bias in direction, dependent and independent variables measured with noise tend to bias estimated coefficients towards the null as its harder to detect signals in such situations – and as the authors report, they do observe that most of their policy coefficients are relatively small and most are not significant.

**Bias Due to Specification?** The authors have estimated time-series models with 2-week and 4-week lags between the policy intensity predictor and the epidemic outcome being modeled. Of note, as the authors themselves acknowledge, the policy variable represents the time when the policy was announced at the national level and not necessarily when it was *actually implemented* in any given jurisdiction. If we consider an outcome like detected cases (or even deaths), then there are naturally lags from when individuals are first infected to when they get tested to when their results are received and their cases recorded (and in the case of death from the time that acute illness occurs to when a person's condition deteriorates and dies).  I would argue that based on the timeline described here, the lags used by the authors are implausibly short. But the question is whether mis-specifying the time-structure of the relationship in the regression can bias outcomes? It is unclear but consider the example in the figure below.

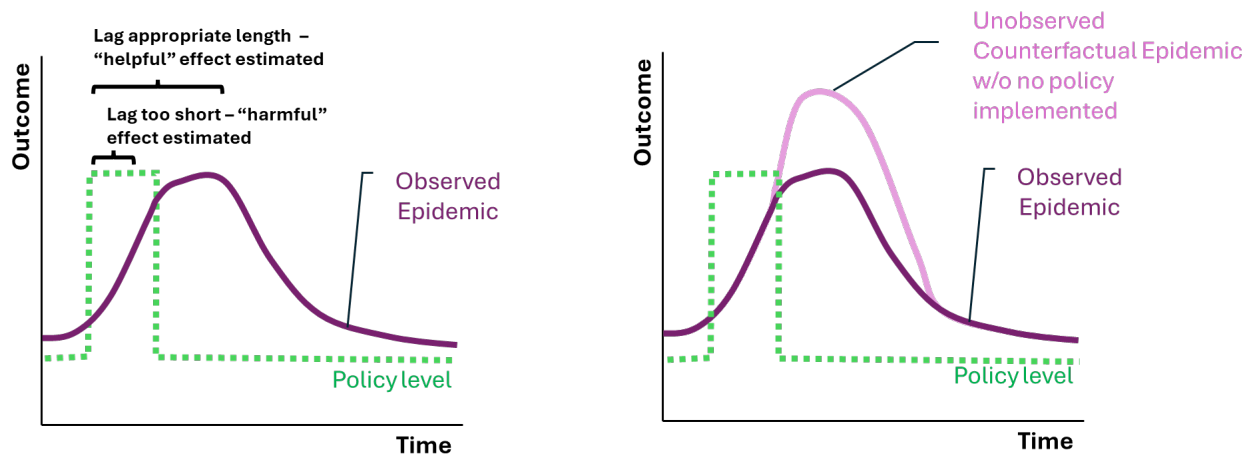

A fixed-effects, short-lag regression analysis would show that when the policy level (green dashed line) is high, the epidemic outcomes are higher/worse and/or rising/worsening, and that when the policy level is lower, the epidemic outcomes are better/declining – a negative regression coefficient for the policy variable or "harmful" in the parlance of the authors' paper. However, if the appropriate, longer lag were used, then the opposite would be shown, and the estimated policy effect would be "helpful".

**Bias Due to Analytic Assumptions Ignoring "Infectiousness"?** To a lesser or greater extent, the time series of countries' outcomes (infections, cases, etc.) are intrinsically linked via "infectiousness" as they describe what the world experienced during a global pandemic caused by a transmissible pathogen. Likewise, for a variety of direct and indirect reasons, the timing and levels of policy responses across countries were also likely linked, as were the populations' behavioral responses to both the evolving epidemics and policies.

However, like many analyses of these pandemic, the authors' implement regression models that do not account for the potential effects of these types of "infectiousness" in estimating policy effects, essentially the relationship between each country's epidemic curves and its policies is assumed to be unaffected by the epidemics and policies of other countries. The units of observation in the authors' analyses are countries, and while it is certainly likely that, especially for larger and for more isolated countries, once an epidemic gets going, within-country transmission is almost certainly the dominant driving force. But earlier in an epidemic when there are fewer infections/cases or during epidemic troughs of infections/cases, imported infections/cases from other countries can contribute meaningfully to the patterns of subsequent epidemic (9).

Imagine that one country implements a policy earlier in the epidemic that keeps its number of cases/infections low for a period of time, but a neighboring country or another country that is not neighboring but has a lot of travel/trade with the country does not implement the policy and has a lot of infections/cases, some of which are imported and drive a larger epidemic in the first country. A regression analysis making assumptions like the one conducted in the authors' paper and other analyses that use such approaches might well estimate that the effect of the policy in the first country that is attenuated (biased towards the null relative to the same policy intensity implemented in a situation without imported cases from other countries) or perhaps even negative (if the number of imported cases and those they transmit too is large enough).

*Brief Discussion of Potential Remedies*

Based on the evaluation presented here, understanding the relationship of policy intensity and epidemic analyses does not require a theory-agnostic, data-driven multiverse set of regression analyses. Instead, it should include:

1) The specification of causal models (including their lag structures) based on biological, mechanistic, behavioral, and other factors that other evidence and theory suggest are at play
2) Use of data at the appropriate scale to avoid/minimize aggregation biases and of appropriate quality to reflect what the analyses are focused on (e.g., not when policies are announced but when they are implemented).
3) Estimation of specified causal models employing techniques that explicitly incorporate "infectiousness" across units of analysis and over time into their specification and estimation.

*Technical Appendix*

The following R code generates the data for the main figure in the manuscript which shows the irrelevance of models that are confounded/not causally correct for assessing the effect of an intervention on an epidemic outcome

```r
library(dplyr)
library(dtplyr)
library(ggplot2)

G_N_SAMPLES        <- 2500
G_GOOD             <- 0
G_NEIGHBORHOOD     <- 0.02
G_BIAS             <- -0.2

v_distance   <- runif(G_N_SAMPLES, min = G_GOOD - 0.5, max = G_GOOD + 0.5)
v_score_bad  <- rnorm(G_N_SAMPLES, mean = G_BIAS, sd = 0.10)
v_score_good <- rnorm(G_N_SAMPLES, mean = 0.3, sd = 0.20)
v_score      <- rep(0, G_N_SAMPLES)
v_dcat       <- rep("No Cat", G_N_SAMPLES)
v_benefit    <- rep("No Benefit", G_N_SAMPLES)

for(i in 1:G_N_SAMPLES) {
  v_score[i] <- ifelse(v_distance[i] > (G_GOOD - G_NEIGHBORHOOD) & v_distance[i] < (G_GOOD + G_NEIGHBORHOOD),
                       v_score_good[i],
                       v_score_bad[i])
  v_dcat[i] <- ifelse(v_distance[i] > (G_GOOD - G_NEIGHBORHOOD) & v_distance[i] < (G_GOOD + G_NEIGHBORHOOD),
                       "Correct Causal Model",
                       "Wrong Causal Model")
  v_benefit[i] <- ifelse(v_score[i]>0, "Helpful", "Harmful")
}

df <- data.frame(distance = v_distance,
                 score    = v_score,
                 mtype    = v_dcat,
                 benefit  = v_benefit) %>%
         mutate(f_good = ifelse(benefit == "Helpful", 1, 0))

mean_wrong    <- mean((df %>% filter(mtype == "Wrong Causal Model"))$score)
mean_right    <- mean((df %>% filter(mtype == "Correct Causal Model"))$score)
mean_overall  <- mean((df)$score)
n_wrong       <- length((df %>% filter(mtype == "Wrong Causal Model"))$f_good)
n_right       <- length((df %>% filter(mtype == "Correct Causal Model"))$f_good)
n_overall     <- length((df)$f_good)
n_good_wrong  <- sum((df %>% filter(mtype == "Wrong Causal Model"))$f_good)
n_good_right  <- sum((df %>% filter(mtype == "Correct Causal Model"))$f_good)
n_good_overall <- sum((df)$f_good)

mean_wrong
mean_right
```

```
mean_overall
n_wrong
n_right
n_overall
n_good_wrong
n_good_right
n_good_overall

n_good_wrong/n_wrong
n_good_right/n_right
n_good_overall/n_overall

ggplot(df, aes(x=distance, y=score, shape = mtype, colour = benefit)) +
  geom_hline(yintercept = 0, size=0.7) +
  geom_vline(xintercept = G_GOOD, size=0.7) +
  geom_vline(xintercept = G_GOOD - G_NEIGHBORHOOD, size=0.7,
             linetype = "dotted") +
  geom_vline(xintercept = G_GOOD + G_NEIGHBORHOOD, size=0.7,
             linetype = "dotted") +
  geom_point(alpha = 0.4) +
  scale_color_brewer(palette = "Set1") +
  geom_hline(yintercept = mean_overall, size = 0.7, color = "darkred") +
  geom_hline(yintercept = mean_right, color = "blue") +
  geom_hline(yintercept = mean_wrong, color = "darkred") +
  theme_bw() +
  ylab("Estimated Intervention Effect") +
  xlab("Distance from Correct Causal Model")
```